\documentclass[aps,prl,twocolumn,showpacs,superscriptaddress]{revtex4}
\usepackage{graphicx}
\begin{document}
\title{Comment on ``Measuring the Orbital Angular Momentum of a
Single Photon''}
\author{Haiqing Wei}
\email[]{dhwei@gazillionbits.com}
%\thanks{}
\affiliation{Gazillion Bits Incorporated, 110 Rose Orchard Way,
San Jose, California 95134} \affiliation{Department of Electrical
\& Computer Engineering, McGill University, Montreal, Canada
H3A-2A7}
\author{Xin Xue}
\affiliation{Gazillion Bits Incorporated, 110 Rose Orchard Way,
San Jose, California 95134}
\begin{abstract}
Optical modes with different orbital angular momentums (OAMs) per
photon may be sorted by Mach-Zehnder interferometers incorporated
with beam rotators, without resorting to OAM mode converters.
\end{abstract}
%42.50.Ct Quantum description of interaction of light and matter;
%related experiments
%42.60.Jf Beam characteristics: profile, intensity, and power;
%spatial pattern formation
%\keywords{angular momentum of photons, mode analysis,
%Laguerre-Gaussian mode.}
\maketitle

Presented in a recent letter \cite{Leach02} is an ingenious method
of sorting spatial modes of photons with different orbital angular
momentum (OAM), which closely resembles, and in a sense
complements a scheme of analyzing optical beams with rectangular
symmetry \cite{Xue01}. The method in \cite{Leach02} employs
special Mach-Zehnder (MZ) interferometers as binary branching
devices, each of which divides a set of input modes with $l\equiv
k~({\rm mod}~2^n)$ into two groups with $l\equiv k~({\rm
mod}~2^{n+1})$ and $l\equiv 2^n+k~({\rm mod}~2^{n+1})$
respectively, where the integer $l$ denotes the single-photon OAM
in units of $\hbar$, $k$ and $n$ are fixed, non-negative integers
specific to the individual interferometer. In the Letter, the same
condition is assumed implicitly for all the MZ interferometers,
that one arm maintains zero phase (or integral multiples of
$2\pi$) for all the modes, while the other arm rotates the beam by
an angle $\alpha$ so to induce an OAM-dependent phase shift
$l\alpha$. The assumption proves rather restrictive and
responsible for the necessity of OAM mode converters, which result
in optical loss and increase significantly the complexity of the
optical setup. The loss of light could impose a serious limitation
to quantum optical experiments involving single photons.

\begin{figure}[h]
\centerline{\scalebox{.5}{\includegraphics{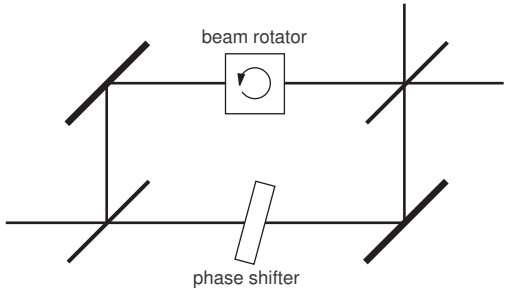}}}
\caption{\label{ourMZ} A modified Mach-Zehnder interferometer.}
\end{figure}

The use of OAM mode converters may be avoided by relaxing the
aforementioned restriction on the MZ interferometers, namely, by
incorporating an adjustable phase shifter in a modified MZ
interferometer as depicted in Fig.\ref{ourMZ}. In practice, there
is always a mechanism of phase adjustment in setting up an optical
interferometer. An example implementation is a thin glass film
making an adjustable angle to the beam axis. When OAM modes with
$l\equiv k~({\rm mod}~2^n)$ are input into the modified MZ
interferometer, the beam is rotated by $\alpha=\pi/2^n$ in the
upper arm such that a mode with OAM $l$ acquires a phase
$l\alpha=l\pi/2^n$, while the phase shifter in the lower arm is
tuned to induce a fixed phase shift $k\alpha=k\pi/2^n$ to all the
OAM modes. It is easily seen that the modified MZ interferometer
segregates the OAM modes into two groups with $l\equiv k~({\rm
mod}~2^{n+1})$ and $l\equiv 2^n+k~({\rm mod}~2^{n+1})$
respectively. Such modified MZ interferometers may be used as
binary branching devices to construct an OAM mode sorter, without
resorting to OAM mode converters.

The OAM mode sorter is in striking similarity to the
Hermite-Gaussian (HG) mode analyzer using MZ interferometers
incorporated with fractional Fourier transformers (FRFTs)
\cite{Xue01}. FRFT-incorporated MZ interferometers can even be
used after an OAM mode sorter to lift the mode degeneracy due to
the radial degree of freedom, so that orthogonal Laguerre-Gaussian
(LG) modes can be sorted completely. Moreover, a complete HG mode
sorter may be utilized just as a complete LG mode analyzer with
the help of HG $\rightleftharpoons$ LG mode converters
\cite{Beijersbergen93}. Finally, it may be noted that the
increased information capacity using higher order OAM
\cite{Leach02,Molina-Terriza02} or HG modes is just the result of
an enlarged spatial channel \cite{Miller98}. To achieve the higher
capacity, the spatial channel needs to support optical beams with
larger sizes than the fundamental mode.

%\begin{acknowledgments}
%Thank ...
%\end{acknowledgments}


\end{document}